\documentclass[pdflatex,sn-mathphys-num]{sn-jnl}% Math and Physical Sciences Numbered Reference Style
\usepackage{graphicx}%
\usepackage{multirow}%
\usepackage{amsmath,amssymb,amsfonts}%
\usepackage{amsthm}%
\usepackage{mathrsfs}%
\usepackage[title]{appendix}%
\usepackage{xcolor}%
\usepackage{textcomp}%
\usepackage{manyfoot}%
\usepackage{booktabs}%
\usepackage{algorithm}%
\usepackage{algorithmicx}%
\usepackage{algpseudocode}%
\usepackage{listings}%
\usepackage{placeins}
\usepackage{subcaption}
\theoremstyle{thmstyleone}%
\theoremstyle{thmstyletwo}%

\theoremstyle{thmstylethree}%

\begin{document}

\title[Analytical Solutions of the Klein–Gordon Oscillator in Som–Raychaudhuri Space–Time]{Analytical Solutions of the Generalized Klein--Gordon Oscillator in Som--Raychaudhuri Space--Time via the extended Nikiforov--Uvarov Method }


\author*[1]{\fnm{Hale} \sur{Karayer}}\email{hale.karayer@klu.edu.tr}

\author[2]{\fnm{Tolga} \sur{Çelik}}\email{tolga.celik.klu@gmail.com}
\author[3]{\fnm{Doğan} \sur{Demirhan}}\email{dogan.demirhan@ege.edu.tr}

\affil*[1]{\orgdiv{Department of Physics}, \orgname{Kırklareli University},
\city{Kırklareli}, \postcode{39100},\country{Turkey}}

\affil[2]{\orgdiv{Department of Physics}, \orgname{Kırklareli University}, %\orgaddress{\street{Street},
\city{Kırklareli}, \postcode{39100}, %\state{State},
\country{Turkey}}

\affil[3]{\orgdiv{Department of Physics}, \orgname{Ege University}, %\orgaddress{\street{Street},
\city{İzmir}, \postcode{35100}, %\state{State},
\country{Turkey}}

%\affil[3]{\orgdiv{Department}, \orgname{Organization}, \orgaddress{\street{Street}, \city{City}, \postcode{610101}, \state{State}, \country{Country}}}


\abstract{In this study, we present an analytical approach based on the extended Nikiforov–Uvarov method to solve the generalized Klein–Gordon oscillator in the presence of a uniform magnetic field within the Som–Raychaudhuri space–time. Exact eigenstate solutions are obtained for two distinct potential models, namely the linear potential and the Cornell potential. The corresponding energy eigenvalues are determined, and the associated eigenfunctions are analyzed and illustrated graphically.}




\maketitle

\section{Introduction}\label{sec1}
Topological defects arising from phase transitions in early-universe scenarios, particularly cosmic strings, play a significant role in both cosmology and field theory. Within the framework of General Relativity, the Gödel space–time metric in the presence of a cosmic string emerges as one of the earliest cosmological solutions describing rotating matter distributions. Owing to its compact structure, this metric provides a suitable gravitational background for the analytical investigation of a wide range of physical and mathematical systems involving rotation and potential causality violations. Moreover, it has been shown that all space–time homogeneous Gödel-type metrics, characterized by a vorticity parameter and a specific parameter $\mu$, can be transformed into polar coordinates \cite{omar}.

In relativistic quantum mechanics, the Klein–Gordon oscillator represents one of the most successful quantum oscillator models, since it can be reduced to the Schrödinger oscillator in the non-relativistic limit. This oscillator was first proposed by Bruce and Minning, inspired by the Dirac oscillator \cite{bruce}. In recent years, this relativistic quantum oscillator model has been extensively studied under various conditions, including confining central potentials, relativistic Landau quantization effects, non-inertial frames, scenarios involving Lorentz symmetry violation, cosmic string space–times, torsioned space–times, Kaluza–Klein theories, the Som–Raychaudhuri space–time, and anti-de Sitter space–time \cite{soares,leite}.

It is well known that, in the limit where the curvature parameter vanishes, Gödel-type space–times reduce to the Som–Raychaudhuri space–time. The solution introduced by Som and Raychaudhuri, within the context of the Einstein–Maxwell equations, belongs to the class of stationary, cylindrically symmetric space–times and describes a rigidly rotating charged dust distribution \cite{omar}. In this configuration, the Lorentz force vanishes everywhere, while the charge-to-mass density ratio can take arbitrary values.
The Som–Raychaudhuri space–time belongs to the class of flat Gödel-type solutions and has been the subject of numerous investigations addressing its physical properties. Studies in this background include analyses of rotating homogeneous structures, the relativistic quantum dynamics of spin-0 particles in nontrivial topological settings, the behavior of the Klein–Gordon oscillator in cosmological scenarios, and the confinement of scalar particles both under linear potentials and in topologically simple flat Gödel-type space–times \cite{Bouzenada}.

In addition, the quantum dynamics of scalar and spin-1/2 particles in Gödel-type space–times with different curvature parameters have been investigated in detail. These studies reveal that the corresponding energy spectra exhibit notable similarities with the Landau problem in flat, spherical, and hyperbolic geometries. In particular, analyses performed in the Som–Raychaudhuri background provide a comparative framework between the quantum dynamics of scalar particles and the Landau levels in flat space.
Furthermore, the quantum dynamics of spin-1/2 particles (Dirac fermions) in the Som–Raychaudhuri space–time with torsion and a cosmic string have been investigated, showing that the degeneracy of relativistic energy levels is lifted and that the corresponding eigenfunctions are directly influenced by the presence of topological defects. Similar approaches have been employed to study Weyl fermions, scalar particles in Gödel-type space–times with cosmic strings, as well as the Fermi field and the Dirac oscillator in the Som–Raychaudhuri background \cite{Bouzenada}.

Studying the generalized Klein–Gordon oscillator in the Som–Raychaudhuri space–time under external magnetic fields and interaction potentials is an important and nontrivial problem in relativistic quantum mechanics \cite{bruce,bakke1,ahmed1,ahmed2}. In such systems, the resulting radial equations are typically reduced to Heun-type differential equations, whose solutions are often obtained only for special cases or under series truncation conditions \cite{soares,olivera,zhong,araujo}.

The main objective of this work is to systematically apply the extended Nikiforov–Uvarov (NU) method to relativistic wave equations defined in curved space–time and to develop unified algebraic solution framework. The extended Nikiforov--Uvarov method is constructed by extending the polynomial degrees appearing in the basic equation of the standard Nikiforov--Uvarov formalism \cite{Nikiforov,berkdemir}. This generalization enlarges the applicability of the method and enables the analytical treatment of a wider class of differential equations, including those associated with Heun-type functions \cite{heun} and relativistic wave equations \cite{hale2}. Within this approach, we focus on deriving analytical solutions of the Klein–Gordon oscillator in the Som–Raychaudhuri space–time and to investigate in a systematic manner the effects of the space–time geometry on the energy spectrum, eigenfunctions, and physical observables.

In contrast to conventional treatments based on ansatz-driven solutions, the proposed method allows the direct solution of the governing differential equations through the extended NU formalism, yielding explicit expressions for both the energy spectra and eigenfunctions. As a result, the quantization conditions are obtained in a purely algebraic form, without the need for series truncation. Furthermore, the effects of topological defect parameters, rotational properties of the space–time, and magnetic fluxes become explicitly traceable within the spectral structure.

The aim of this study is to provide a systematic analytical framework for relativistic quantum systems in curved space–time, providing an alternative to non-unified  and problem-specific methods commonly used in the literature. The extended NU-based framework is also expected to be directly applicable to a broader class of problems, including those arising from Kaluza–Klein reductions and effective Schrödinger-type equations induced by extra-dimensional theories.
The organization of this paper is as follows. In Sec. 2, we present an overview of the Klein–Gordon oscillator in the Som–Raychaudhuri space–time. In Sec. 3, we introduce the fundamental formalism of the extended NU method. In Sec. 4, we derive the analytical solutions of the Klein–Gordon oscillator for both linear and Cornell-type potentials in the presence of a uniform magnetic field. Finally, our concluding remarks are provided in Sec. 5.
\section{Klein–Gordon Oscillator in the Som–Raychaudhuri Space–time}
We start by describing space-time using cylindrical coordinates in the Som Raychaudhuri limit. In this geometry, the line element is written as
\begin{equation}
ds^{2}
=
-\left(dt+\alpha \Omega r^{2} d\phi\right)^{2}
+\alpha^{2}r^{2}d\phi^{2}
+dr^{2}
+dz^{2}.
\end{equation}
where the coordinates satisfy
$r \geq 0,  0 \leq \phi \leq 2\pi,  -\infty < z < \infty.$ \cite{zhong}

Here, the parameter \(\alpha\) represents the angular defect of the space-time and is directly related to the linear mass density \(\eta\) through the relation
\begin{equation}
\alpha = 1 - 4\eta.
\end{equation}
The parameter $\Omega$ denotes the vorticity (rotation) parameter of the space--time.
For this background, the inverse metric tensor is written as;
\[
g^{\mu\nu}=
\begin{pmatrix}
r^{2}\Omega^{2}-1 & 0 & \dfrac{-\Omega}{\alpha} & 0\\
0 & 1 & 0 & 0\\
\dfrac{-\Omega}{\alpha} & 0 & \dfrac{1}{\alpha^{2}r^{2}} & 0\\
0 & 0 & 0 & 1
\end{pmatrix}.
\] 

The dynamics of relativistic quantum particles in curved space-time can be described by the Klein--Gordon equation, which we adopt throughout this work \cite{zhong}.
\begin{equation}
\frac{1}{\sqrt{-g}}
\left(D_{\mu}+M\omega X_{\mu}\right)
\left[
\sqrt{-g}\,g^{\mu\nu}
\left(D_{\nu}-M\omega X_{\nu}\right)
\right]\Psi
=
\left(M+S(r)\right)^{2}\Psi .
\end{equation}
where \(g\) is the determinant of the metric tensor, with
\begin{equation}
\sqrt{-g} = \alpha r.
\end{equation}

In this equation, \(M\) denotes the particle mass, while \(S(r)\) is a scalar potential. The operator \(D_{\mu}\) is the covariant derivative that incorporates electromagnetic interactions through the vector potential.

Physically, scalar potentials are generally associated with static field configurations, whereas vector potentials are related to dynamic electromagnetic fields. Therefore, the scalar potential can be interpreted as an additional contribution to the mass energy of the particle.
Within the minimal coupling scheme, the covariant derivative is defined as
\begin{equation}
D_{\mu} = \partial_{\mu} - ieA_{\mu},
\end{equation}
where \(e\) is the electric charge and \(A_{\mu}\) is the electromagnetic four-potential. In this work, the four-potential is chosen in the form
$A_{\mu} = (0,0,A_{\phi},0).$
Substituting this choice into the Klein--Gordon equation leads to the differential equation ($c = \hbar = 1.$);
\begin{eqnarray}\label{gkg}
[
-\frac{\partial^{2}}{\partial t^{2}}
+
\left(
r\Omega\frac{\partial}{\partial t}
-
\frac{1}{\alpha r}
\frac{\partial}{\partial \phi}
-ieA_{\phi}
\right)^{2}
+
\frac{\partial^{2}}{\partial r^{2}}
+
\frac{1}{r}\frac{\partial}{\partial r}
-\frac{M\omega}{r}f(r)
-M\omega f'(r) \nonumber \\
-M^{2}\omega^{2}f^{2}(r)
+
\frac{\partial^{2}}{\partial z^{2}}
-
\left(M+S(r)\right)^{2}
]\Psi=0.
\end{eqnarray}
where the momentum operator undergoes the transformation,
$p_{\mu} \rightarrow p_{\mu}+iM\omega X_{\mu},$
for
$X_{\mu}=(0,f(r),0,0),$
with \(f(r)\) being an arbitrary radial function and \(\omega\) denoting the oscillator frequency \cite{mirza}.

Eq.~(\ref{gkg}) is the  generalized form of the Klein--Gordon equation which contains additional terms associated with the oscillator coupling, making it possible to study relativistic oscillator dynamics in curved space-time \cite{lutfu}.

The azimuthal component of the vector potential is taken as
\begin{equation}
A_{\phi} = -\frac{\alpha B_{0}}{2}r^{2} + \frac{\Phi_{B}}{2\pi}.
\end{equation}
where \(\Phi_{B}\) denotes an internal quantum magnetic flux, assumed to be constant, and \(B_{0}\) is the strength of the external magnetic field.

To solve the wave equation, we assume that the wave function can be separated into temporal, angular, axial, and radial parts. Thus, we use the ansatz
\begin{equation}\label{psisprt}
\Psi(t,r,\phi,z)=e^{i(-Et+\ell\phi+kz)}s(r),
\end{equation}
where \(k\) is the wave number along the \(z\)-direction, \(\ell = 0,\pm1,\pm2,\dots\) is the angular momentum quantum number, \(E_{n,\ell}\) represents the energy eigenvalue and \(s_{n,\ell}(r)\) is the radial wave function. Finally, substituting the chosen potentials and the wave-function ansatz into the generalized Klein--Gordon equation leads to the radial differential equation. This equation governs the radial behavior of the relativistic quantum particle under the combined effects of curved space-time, oscillator interaction, magnetic field, scalar potential, and topological defects.

\section{Formulation of the Extended Nikiforov–Uvarov Framework}

The extended NU approach is a powerful analytical technique for solving second-order differential equations of a generalized type. The method is based on the transformation of the wave equation into the following canonical form.
\begin{equation}
\label{eq:ex_nu_new}
\frac{d^{2}\psi(z)}{dz^{2}}+\frac{\tilde{\tau}_{e}(z)}{\sigma_{e}(z)}\frac{d\psi(z)}{dz}+\frac{\tilde{\sigma}_{e}(z)}{\sigma_{e}^{2}(z)}\psi(z)=0,
\end{equation}
where $\tilde{\tau}_{e}(z)$, $\sigma_{e}(z)$, and $\tilde{\sigma}_{e}(z)$ denote polynomials whose maximum degrees are two, three, and four, respectively. The subscript $e$ indicates the extended version of the method \cite{hale2,hale3,hale4}.
For a physically meaningful solution, the wavefunction $\psi(z)$ must satisfy standard boundary conditions, namely being single-valued, continuous, and vanishing asymptotically.
Within this formalism, the determination of the energy spectrum and corresponding eigenfunctions is based on a set of auxiliary relations. The function $h(z)$, associated with the eigenvalue problem, is expressed as
\begin{equation}
\label{hn_new}
h(z)=h_{n}(z)=-\frac{n}{2}\frac{d\tau_{e}(z)}{dz}-\frac{n(n-1)}{6}\frac{d^{2}\sigma_{e}(z)}{dz^{2}}+C_{n},
\end{equation}
where $n$ is a non-negative integer and $C_n$ is a constant.
The logarithmic derivative of the function $\phi(z)$ is given by
\begin{equation}
\label{phie_new}
\frac{d}{dz}\ln \phi(z)=\frac{\pi_{e}(z)}{\sigma_{e}(z)},
\end{equation}
which allows the construction of the total wavefunction in a factorized form.
Polynomial solutions are obtained from the differential equation
\begin{equation}
\label{asl_new}
\sigma_{e}(z)\frac{d^{2}y(z)}{dz^{2}}+\tau_{e}(z)\frac{dy(z)}{dz}+h(z)y(z)=0,
\end{equation}
where the function $\tau_{e}(z)$ is defined as
\begin{equation}
\label{eq:ex_tau_new}
\tau_{e}(z)=\tilde{\tau}_{e}(z)+2\pi_{e}(z).
\end{equation}
The function $\pi_{e}(z)$ is determined by the relation
\begin{equation}
\label{eq:ex_pi_new}
\pi_{e}(z)=\frac{\sigma'_{e}(z)-\tilde{\tau}_{e}(z)}{2}\pm \sqrt{\left(\frac{\sigma'_{e}(z)-\tilde{\tau}_{e}(z)}{2}\right)^{2}-\tilde{\sigma}_{e}(z)+{g}(z)\sigma_{e}(z)},
\end{equation}
where $\pi_{e}(z)$ is restricted to be a polynomial of degree at most two. The auxiliary function ${g}(z)$ is introduced as
\begin{equation}
\label{eq:ex_hz_new}
g(z)=h(z)-\frac{d\pi_{e}(z)}{dz}.
\end{equation}
The complete solution of Eq.~(\ref{asl_new}) can be written as $y(z)=y_{n}(z)$, where $y_{n}(z)$ is a polynomial of degree $n$. Accordingly, the total wavefunction takes the form
\begin{equation}
\psi(z)=\phi(z)\,y_{n}(z).
\end{equation}
It is important to note that the functions $h(z)$ and $\tau_{e}(z)$ that arise during the solution procedure are, at most, first- and second-degree polynomials, respectively. The quantization condition is obtained by imposing $h(z)=h_{n}(z)$, from which the energy eigenvalues are established. The corresponding eigenfunctions are then constructed by combining the function $\phi(z)$ with the polynomial solution $y_{n}(z)$ \cite{hale2,hale3,hale4,hale5}.
\section{Analytical Solutions of the Klein Gordon Oscillator by the Extended Nikiforov–Uvarov}
\subsection{The linear potential}
The Klein--Gordon oscillator equation given by Eq.~(\ref{psisprt}) in the presence of a Coulomb-type scalar interaction can be written as
\begin{eqnarray} \label{bbb}
\left[
-\frac{\partial^{2}}{\partial t^{2}}
+
\left(
r\Omega\frac{\partial}{\partial t}
-
\frac{1}{\alpha r}
\frac{\partial}{\partial \phi}
-
ieA_{\phi}
\right)^{2}
+
\frac{\partial^{2}}{\partial r^{2}}
+
\frac{1}{r}\frac{\partial}{\partial r}
\nonumber \right. \\
\left.
-2M\omega
-M^{2}\omega^{2}r^{2}
+
\frac{\partial^{2}}{\partial z^{2}}
-
\left(M+\frac{\lambda}{r}\right)^{2}
\right]\Psi=0,
\end{eqnarray}
where the scalar potential is chosen in the Coulomb form, $S(r)=\frac{\lambda}{r}$, together with the linear potential $f(r)=r$. Substituting Eq.~(\ref{psisprt}) into Eq.~(\ref{bbb}), the radial part of the equation becomes \cite{zhong}
\begin{equation}\label{scomp}
s'' + \frac{1}{r}s' + \frac{1}{r^2}
\left[
a_4 r^2 - a_1 - a_2^2 r^4 - 2a_3 r
\right] s = 0.
\end{equation}
The coefficients appearing in Eq.~(\ref{scomp}) are defined as
\begin{eqnarray}\label{prmtra}
a_1 &=&
\frac{l^2}{\alpha^2}
+\frac{e^2\Phi_B^2}{4\alpha^2\pi^2}
+\lambda^2
-\frac{el\Phi_B}{\alpha^2\pi},
\nonumber \\
a_2^2 &=&
E^2\Omega^2
+e\Omega B_0 E
+\frac{e^2B_0^2}{4}
+M^2\omega^2,
\nonumber \\
a_3 &=&
M\lambda,
\nonumber \\
a_4 &=&
E^2
+\frac{\Omega e\Phi_B}{\alpha\pi}E
-\frac{2\Omega l}{\alpha}E
-\frac{eB_0 l}{\alpha}
+\frac{e^2\Phi_B B_0}{2\pi\alpha}
-2M\omega
-k^2
-M^2.
\end{eqnarray}
Equation~(\ref{scomp}) satisfies the conditions required by the extended NU method given in Eq.~(\ref{eq:ex_nu_new}). By comparing Eq.~(\ref{scomp}) with the basic equation of the method, the corresponding polynomials are identified as
\begin{eqnarray}
\tau_e(r)&=&1,\\
\sigma_e(r)&=&r, \\
\tilde{\sigma}_e(r)
&=&
-a_2^2 r^4
+a_4 r^2
-2a_3 r
-a_1.
\end{eqnarray}
Using these expressions, the possible forms of the polynomial $\pi_e(r)$ are obtained from Eq.~(\ref{eq:ex_pi_new}) as
\begin{equation}\label{aaaa}
\pi_{e1}(r)=a_2 r^2+\sqrt{a_1},
\end{equation}
\begin{equation}
\pi_{e2}(r)=-a_2 r^2-\sqrt{a_1},
\end{equation}
corresponding to
\[
g_1(r)=(a_4+2a_2\sqrt{a_1})r-2a_3.
\]
and
\begin{equation}
\pi_{e3}(r)=-a_2 r^2+\sqrt{a_1},
\end{equation}
\begin{equation}
\pi_{e4}(r)=a_2 r^2-\sqrt{a_1},
\end{equation}
corresponding to
\[
g_2(r)=(a_4-2a_2\sqrt{a_1})r-2a_3,
\]
All possible forms of $\pi_e(r)$ generates a different eigenstate solution. However, in order for the solutions to be physically acceptable, the appropriate forms among the four possible expressions of the polynomial $\pi_e(r)$, where the first derivative of the polynomial $\tau_e(r)$ is negative, must be employed. Only these forms satisfy the condition
$\tau'_e(r)<0$,
which is required for normalizable bound-state solutions. In this context, physically acceptable solutions are obtained for $\pi_{e2}(r)$ and $\pi_{e3}(r)$.To avoid repetitive calculations, only the solution associated with $\pi_{e2}(r)$ is presented in detail, while the remaining solution are given directly.
From Eqs.~(\ref{eq:ex_hz_new}), (\ref{eq:ex_tau_new}), and (\ref{hn_new}), the following auxiliary polynomials are obtained:
\begin{equation}
h(r)=a_4+2a_2(-1+\sqrt{a_1})r-2a_3,
\end{equation}
\begin{equation}
\tau_e(r)
=
-2a_2 r^2 - 2\sqrt{a_1}+1,
\end{equation}
and
\begin{equation}
2na_2 r + C_n
=
(a_4+2a_2(-1+\sqrt{a_1}))r-2a_3.
\end{equation}
The above relation yields both the energy condition and the integration constant:
\begin{equation}\label{egn1}
2n
=
\frac{a_4}{a_2}
-2
+2\sqrt{a_1},
\end{equation}
\begin{equation}
C_n=-2a_3.
\end{equation}
The function $\Phi(r)$ appearing in the eigenfunction solution is obtained from Eq.~(\ref{phie_new}) as;
\begin{equation}
\Phi(r)
=
e^{-\frac{a_2 r^2}{2}}
r^{-\sqrt{a_1}}.
\end{equation}
To determine the polynomial part of the wave function, we start from Eq.~(\ref{asl_new}) and obtain
\begin{equation}\label{eigenfunctrho}
\rho y''(\rho)
+
\left(
1-2\sqrt{a_1}-2\rho^2
\right)
y'(\rho)
+
\left(
\frac{a_4}{a_2}
-2
+2\sqrt{a_1}
-\frac{2a_3}{\sqrt{a_2}}
\right)
y(\rho)
=0,
\end{equation}
where the new variable is defined as $\rho=\sqrt{a_2}r$.
Equation~(\ref{eigenfunctrho}) has the same structure as the biconfluent Heun equation
\begin{equation}
x y''
+
(1+\alpha+\beta x-2x^2)y'
+
[\gamma-\alpha-2-\delta(1+\alpha+\beta)]y
=0,
\end{equation}
whose polynomial solutions are represented by
$N(\alpha,\beta,\gamma,\delta,y)$ provided that the condition
\[
\gamma-\alpha-2=2n
\]
is satisfied \cite{heun,caruso,ishkhanyan}.
Comparing Eq.~(\ref{eigenfunctrho}) with the standard biconfluent Heun form gives
\begin{equation}
\alpha=-2\sqrt{a_1},
\qquad
\beta=0,
\qquad
\gamma=\frac{a_4}{a_2},
\qquad
\delta=\frac{4a_3}{\sqrt{a_2}}.
\end{equation}
Since the condition
\[
\gamma-\alpha-2=2n
\]
is fully consistent with the eigenvalue relation given in Eq.~(\ref{egn1}), Eq.~(\ref{eigenfunctrho}) admits polynomial solutions of the form
\[
N\left(
-2\sqrt{a_1},
0,
\frac{a_4}{a_2},
\frac{4a_3}{\sqrt{a_2}},
\sqrt{a_2}r
\right).
\]
As a result, the complete radial eigenfunction can be expressed as
\begin{equation}\label{36}
\psi(r)
=
e^{-\frac{a_2 r^2}{2}}
r^{-\sqrt{a_1}}
N\left(
-2\sqrt{a_1},
0,
\frac{a_4}{a_2},
\frac{4a_3}{\sqrt{a_2}},
\sqrt{a_2}r
\right).
\end{equation}

The eigensatate solution that is produced by $\pi_{e3}(r)$ in Eq.~(\ref{aaaa}) becomes;
\begin{equation}\label{egn2}
2n=\frac{a_4}{a_2}-2-2\sqrt{a_1},
\end{equation}
\begin{equation}
C_n=-2a_3.
\end{equation}
\begin{equation}\label{39}
\psi(r)=e^{-\frac{a_2 r^2}{2}}\, r^{\sqrt{a_1}}
\, N\left(2\sqrt{a_1},0,\frac{a_4}{a_2},
\frac{4a_3}{\sqrt{a_2}},\sqrt{a_2}\,r\right).
\end{equation}

By inserting the parameters defined in Eq.~(\ref{prmtra}) into the eigenvalue relation given in Eq.~(\ref{egn2}) is obtained as follows:
\begin{eqnarray}
\left(
2n+2
+
2\sqrt{
\frac{\breve{l}_{ef}^{\,2}}{\alpha^{2}}
+\lambda^{2}
}
\right)
\sqrt{
E^{2}\Omega^{2}
+e\Omega B_{0}E
+\frac{e^{2}B_{0}^{2}}{4}
+M^{2}\omega^{2}
}
\nonumber \\
=
E^{2}
-\breve{l}_{ef}
\left(
\frac{2\Omega E}{\alpha}
+\frac{eB_{0}}{\alpha}
\right)
-2M\omega
-k^{2}
-M^{2}.
\end{eqnarray}
where $
\breve{l}_{ef}=l-\frac{e\Phi_{B}}{2\pi}.
$ The effective angular momentum modifies the energy spectrum, leading to an effect analogous to the Aharonov--Bohm effect \cite{aharonov, chen}. Table~\ref{tab:linear_energy} presents the positive energy eigenvalues of the system under the linear potential for different values of the topological parameter $\alpha$. The numerical calculations are performed for $M=e=\Phi_B=\omega=\lambda=B_0=\Omega=k=1$, while the allowed quantum numbers satisfy $n=0,1,2,\ldots$ and $l=0,1,\ldots,n-1$.

\begin{table*}[t]
\caption{Positive energy eigenvalues for the linear potential for different values of the topological parameter $\alpha$. The numerical results are obtained for $M=e=\Phi_B=\omega=\lambda=B_0=\Omega=k=1$.}
\label{tab:linear_energy}
\begin{tabular*}{\textwidth}{@{\extracolsep\fill}ccccc}
\toprule
$n$ & $l$ & $\alpha=0.3$ & $\alpha=0.5$ & $\alpha=0.8$ \\
\midrule
0 & 0 & 4.53166 & 4.7502 & 4.90384 \\
1 & 0 & 6.31984 & 6.55478 & 6.71848 \\
2 & 0 & 8.18911 & 8.43266 & 8.60169 \\
2 & 1 & 18.2741 & 14.9574 & 11.8368 \\
3 & 0 & 10.1022 & 10.3507 & 10.5227 \\
3 & 1 & 20.2542 & 16.0252 & 13.7927 \\
3 & 2 & 33.3263 & 23.6621 & 18.3374 \\
4 & 0 & 12.0409 & 12.2923 & 12.4663 \\
4 & 1 & 22.2377 & 17.9995 & 15.7587 \\
4 & 2 & 35.3199 & 25.6498 & 20.3175 \\
4 & 3 & 48.5657 & 33.5210 & 25.1399 \\
5 & 0 & 13.9956 & 14.2489 & 14.4241 \\
5 & 1 & 24.2237 & 19.9786 & 17.7317 \\
5 & 2 & 37.3142 & 27.6392 & 22.3009 \\
5 & 3 & 50.5266 & 35.5146 & 27.1289 \\
5 & 4 & 63.8521 & 43.4488 & 32.0359 \\
\botrule
\end{tabular*}
\end{table*}

\subsection{The Cornell potential}
The Klein--Gordon oscillator equation in the presence of a Cornell potential which describes behavior of quarks, can be written as
\begin{eqnarray}\label{kkk}
&&\left[
-\frac{\partial^{2}}{\partial t^{2}}
+
\left(
r\Omega\frac{\partial}{\partial t}
-
\frac{1}{\alpha r}\frac{\partial}{\partial \phi}
-
ieA_{\phi}
\right)^{2}
+
\frac{\partial^{2}}{\partial r^{2}}
+
\frac{1}{r}\frac{\partial}{\partial r}
-2M\omega\xi_{1}
-M^{2}\omega^{2}\xi_{1}^{2}r^{2}
\right. \nonumber \\
&&\left.
-2M^{2}\omega^{2}\xi_{1}\xi_{2}
-\frac{M^{2}\omega^{2}\xi_{2}^{2}}{r^{2}}
+
\frac{\partial^{2}}{\partial z^{2}}
-
\left(
M+\frac{\lambda}{r}
\right)^{2}
\right]
\Psi(t,r,\phi,z)=0.
\end{eqnarray}
where the scalar potential is Coulomb form, together with the linear potential $f(r)=\xi_{1}r+\frac{\xi_{2}}{r}$ \cite{zhong}. Substituting Eq.~(\ref{psisprt}) into Eq.~(\ref{kkk}), the radial part of the equation becomes;
\begin{equation}
s''+\frac{1}{r}s'
+
\left(
b_{4}
-\frac{b_{1}}{r^{2}}
-b_{2}^{2}r^{2}
-\frac{2b_{3}}{r}
\right)s=0,
\end{equation}
where \(b_{1}, b_{2}, b_{3}\), and \(b_{4}\) are newly defined parameters given by
\begin{eqnarray}\label{prmtrb}
b_{1}
&=&
M^{2}\omega^{2}\xi_{2}^{2}
+\frac{l^{2}}{\alpha^{2}}
+\frac{e^{2}\Phi_{B}^{2}}{4\alpha^{2}\pi^{2}}
+\lambda^{2}
-\frac{el\Phi_{B}}{\alpha^{2}\pi},
\nonumber \\
b_{2}^{2}
&=&
E^{2}\Omega^{2}
+e\Omega B_{0}E
+M^{2}\omega^{2}\xi_{1}^{2}
+\frac{e^{2}B_{0}^{2}}{4},
\nonumber \\
b_{3}
&=&
M\lambda,
\nonumber \\
b_{4}
&=&
E^{2}
+\frac{\Omega e\Phi_{B}}{\alpha\pi}E
-\frac{2\Omega l}{\alpha}E
-2M\omega\xi_{1}
-2M^{2}\omega^{2}\xi_{1}\xi_{2}
\nonumber \\
&&
-\frac{eB_{0}l}{\alpha}
+\frac{e^{2}\Phi_{B}B_{0}}{2\pi\alpha}
-k^{2}
-M^{2}.
\end{eqnarray}
Since this equation satisfies the basic equation of the extended NU method, it can be solved within the framework of this method. All possible values of the polynomials $\pi_{e}(r)$ can be listed as follows;
\begin{eqnarray}
\pi_{e1}(r) &=& b_2 r^2 + \sqrt{b_1},\\
\pi_{e2}(r) &=& -b_2 r^2 - \sqrt{b_1}
\end{eqnarray}
for $g_1(r) = (b_4 + 2b_2 \sqrt{b_1})r - 2b_3$
\begin{eqnarray}
\pi_{e3}(r) &=& -b_2 r^2 + \sqrt{b_1},\\
\pi_{e4}(r) &=& b_2 r^2 - \sqrt{b_1}
\end{eqnarray}
for $g_2(r) = (b_4 - 2b_2 \sqrt{b_1})r - 2b_3 $.

The method generates different eigenstate solutions for each of these polynomials. Since the solution procedure for the linear potential has already been presented above, only the complete set of solutions corresponding to the physically acceptable states is given here. The resulting eigenfunctions are expressed in terms of biconfluent Heun polynomials \cite{heun,caruso,ishkhanyan}.\\
For $\pi_{e2}(r)$;
\begin{eqnarray}\label{egn3}
2n&=&\frac{b_{4}}{b_{2}}-2+2\sqrt{b_{1}}\\
C_n &=& -2b_3 \\
\psi(r) &=& e^{\frac{-b_2 r^2}{2}} r^{-\sqrt{b_1}} N\left(-2\sqrt{b_1}, 0, \frac{b_4}{b_2}, \frac{4b_3}{\sqrt{b_2}}, \sqrt{b_2}r\right) \label{50}
\end{eqnarray}
For $\pi_{e3}(r)$;
\begin{eqnarray}\label{egn4}
2n &=& \frac{b_4}{b_2} - 2 - 2\sqrt{b_1} \\
C_n &=& -2b_3 \\
\psi(r) &=& e^{\frac{-b_2 r^2}{2}} r^{\sqrt{b_1}} N\left(2\sqrt{b_1}, 0, \frac{b_4}{b_2}, \frac{4b_3}{\sqrt{b_2}}, \sqrt{b_2}r\right) \label{53}
\end{eqnarray}
By substituting the parameters introduced in Eq.~(\ref{prmtrb}) into the eigenvalue equation given in Eq.~(\ref{egn4}), one obtains the following expression for $
\breve{l}_{ef}=l-\frac{e\Phi_{B}}{2\pi}
$:
\begin{eqnarray}
\left[
2n+2
+
2\sqrt{
M^{2}\omega^{2}\xi_{2}^{2}
+\lambda^{2}
+\frac{\breve{l}_{ef}^{\,2}}{\alpha^{2}}
}
\right]
\sqrt{
E^{2}\Omega^{2}
+e\Omega B_{0}E
+M^{2}\omega^{2}\xi_{1}^{2}
+\frac{e^{2}B_{0}^{2}}{4}
}\nonumber \\
=
E^{2}
-2M\omega\xi_{1}
-2M^{2}\omega^{2}\xi_{1}\xi_{2}
-\frac{2\Omega E}{\alpha}\breve{l}_{ef}
-\frac{eB_{0}}{\alpha}\breve{l}_{ef}
-k^{2}
-M^{2}.
\end{eqnarray}
\FloatBarrier
\begin{table*}[htb!]
\caption{Positive energy eigenvalues for the Cornell-type potential for different values of the topological parameter $\alpha$. The numerical results are obtained for $M=e=\Phi_B=\omega=\lambda=B_0=\Omega=k=\xi_2=1$ and $\xi_1=2$.}
\label{tab:cornell_energy}
\begin{tabular*}{\textwidth}{@{\extracolsep\fill}ccccc}
\toprule
$n$ & $l$ & $\alpha=0.3$ & $\alpha=0.5$ & $\alpha=0.8$ \\
\midrule
0 & 0 & 6.14859 & 6.3820 & 6.5364 \\
1 & 0 & 7.82988 & 8.08331 & 8.24957 \\
2 & 0 & 9.60033 & 9.86713 & 10.0413 \\
2 & 1 & 18.9504 & 14.9745 & 12.9554 \\
3 & 0 & 11.4308 & 11.7066 & 11.8862 \\
3 & 1 & 20.9022 & 16.8999 & 14.8583 \\
3 & 2 & 33.6867 & 24.1981 & 19.0719 \\
4 & 0 & 13.3023 & 13.5842 & 13.7675 \\
4 & 1 & 22.8613 & 18.8389 & 16.7806 \\
4 & 2 & 35.6704 & 26.1672 & 21.0233 \\
4 & 3 & 48.8097 & 33.8936 & 25.6707 \\
5 & 0 & 15.2021 & 15.4885 & 15.6743 \\
5 & 1 & 24.8263 & 20.7882 & 18.7174 \\
5 & 2 & 37.6556 & 28.1402 & 22.9821 \\
5 & 3 & 50.8017 & 35.8773 & 27.6427 \\
5 & 4 & 64.0362 & 43.7336 & 32.4486 \\
\botrule
\end{tabular*}
\end{table*}
Table~\ref{tab:cornell_energy} presents the positive energy eigenvalues obtained for the Cornell potential under different values of the topological parameter $\alpha$. The numerical calculations are performed for $M=e=\Phi_B=\omega=\lambda=B_0=\Omega=k=\xi_2=1$ and $\xi_1=2$, while the allowed quantum numbers satisfy $n=0,1,2,\ldots$ and $l=0,1,\ldots,n-1$. It is observed that the energy spectrum is significantly affected by both the angular quantum number $l$ and the topological parameter $\alpha$, reflecting the influence of the cosmic string background on the relativistic bound states.

\section{Results and Discussion}\label{sec12}

In this work, the generalized Klein--Gordon oscillator in the Som--Raychaudhuri space--time was investigated within the framework of the extended Nikiforov--Uvarov method. Exact analytical expressions for both the energy eigenvalues and the corresponding eigenfunctions were obtained for the linear and Cornell-type interaction potentials in the presence of a uniform magnetic field. Unlike conventional approaches based on ansatz assumptions or series truncation procedures, the present method provides a direct algebraic treatment of the governing differential equations, leading to explicit quantization conditions and closed form wave function solutions in terms of biconfluent Heun polynomials. The resulting energy spectra were calculated numerically for different values of the topological parameter $\alpha$, and the corresponding radial wave functions were analyzed to examine the influence of the interaction potentials and quantum numbers on the bound-state structure.

Tables~1 and 2 shows the positive energy eigenvalues obtained for the linear and Cornell-type potentials, respectively. In both cases, the energy spectrum depends strongly on the radial quantum number $n$, the angular quantum number $l$, and the topological parameter $\alpha$. For fixed $l=0$, the energy eigenvalues increase regularly as $n$ increases.
The effect of the topological parameter $\alpha$ is more sensitive to the angular quantum number. The increase in $\alpha$ weakens the influence of the topological defect,
which modifies the effective angular momentum and consequently shifts
the relativistic energy levels. For the states with $l=0$, the energy eigenvalues slightly increase as $\alpha$ increases. However, for states with $l\neq 0$, the opposite trend is observed: the energy decreases when $\alpha$ increases. This indicates that the angular defect of the Som--Raychaudhuri space--time has a stronger influence on states carrying angular momentum. In particular, the term involving the effective angular momentum $\breve{l}_{\rm eff}/\alpha$ plays a dominant role in shifting the relativistic energy levels.
A comparison between Tables~1 and 2 demonstrates that the Cornell-type potential generally produces larger energy eigenvalues than the linear potential for the same quantum numbers and parameter values. This result reflects the stronger confining character of the Cornell-type interaction. The additional inverse-radial contribution in the oscillator function modifies the effective potential and increases the energy required to form the corresponding bound states. Therefore, the Cornell-type interaction generally produces higher energy levels than the purely linear case.

The present results are consistent with the behavior reported in Ref.~\cite{zhong}, where the energy eigenvalues decrease with increasing values of the angular parameter $\alpha$ for states with nonzero angular momentum. However, the numerical analysis presented here further indicates that the response of the spectrum to variations of $\alpha$ depends on the angular quantum number, leading to a different trend for the $l=0$ states.

As illustrated in Fig.~1, all radial wave functions vanish at large distances, indicating that the obtained solutions correspond to physically acceptable bound states. The shape of the wave functions is strongly influenced by both the quantum numbers and the particular eigenstate solutions generated by the extended Nikiforov--Uvarov method. For higher quantum states, the wave functions tend to spread over a larger radial region and exhibit a more pronounced oscillatory behavior, which is a characteristic feature of excited bound states.

The comparison between the linear and Cornell-type potentials confirms that the form of the interaction potential has a noticeable impact on the spatial behavior of the particle. In particular, the additional inverse-radial term contained in the Cornell-type interaction modifies the effective confinement and produces radial profiles that differ from those obtained in the linear-potential case. This demonstrates that the localization properties of the bound states are sensitive to the underlying interaction mechanism.

Moreover, the differences observed among panels (a)--(d) indicate that the admissible eigenstate solutions obtained within the extended Nikiforov--Uvarov framework can display distinct radial characteristics, even though they are derived from the same relativistic wave equation. These consequences display that the extended Nikiforov--Uvarov method provides not only exact analytical expressions for the energy spectrum but also a useful framework for examining how interaction potentials, topological defects, and external fields affect the spatial structure of relativistic bound states in curved space--time.

The success of the method in deriving exact energy spectra and closed-form eigenfunctions without relying on ansatz-based assumptions suggests that it can be applied to a broad range of relativistic wave equations with different potentials, topological defects, external fields, and symmetry structures. In this sense, the present work constitutes a step toward establishing a systematic algebraic framework for the analytical treatment of relativistic wave equations in nontrivial geometrical backgrounds.

\begin{figure}[htbp]
\centering

\begin{subfigure}[t]{0.48\textwidth}
\centering
\includegraphics[
width=\linewidth,
trim={2cm 2cm 2cm 2cm},
clip
]{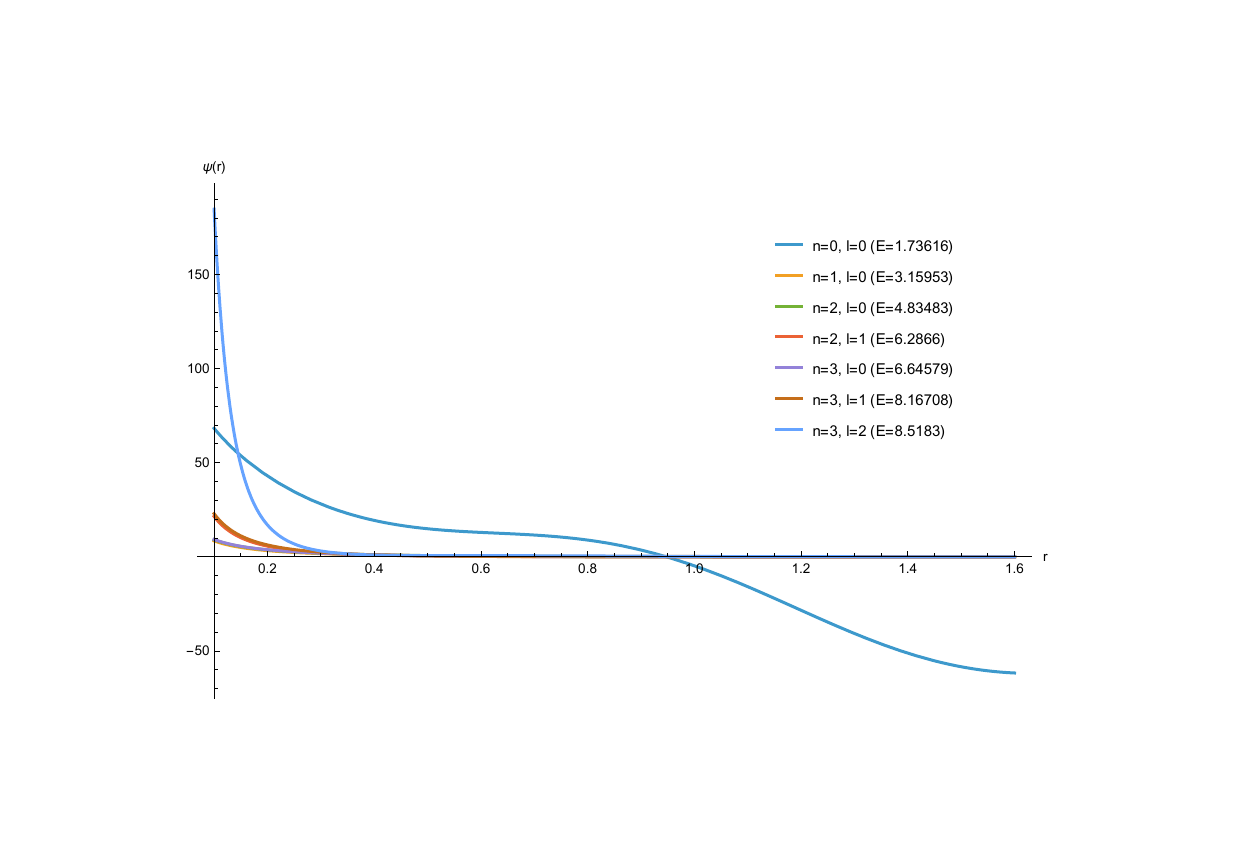}
\caption{Radial wave function in Eq.~(\ref{36})}
\label{fig:linear36}
\end{subfigure}
\hfill
\begin{subfigure}[t]{0.48\textwidth}
\centering
\includegraphics[
width=\linewidth,
trim={2cm 2cm 2cm 2cm},
clip
]{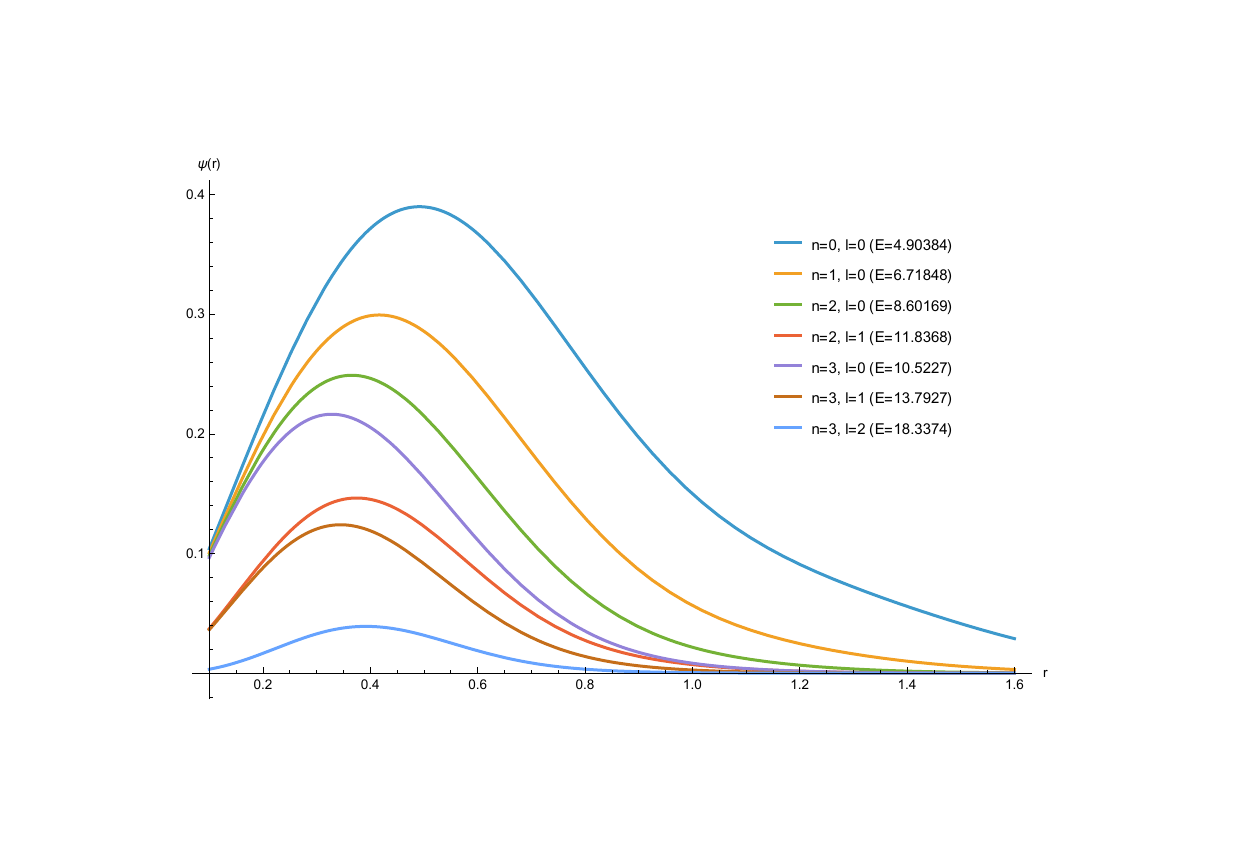}
\caption{Radial wave function in Eq.~(\ref{39})}
\label{fig:linear39}
\end{subfigure}

\vspace{-0.3cm}

\begin{subfigure}[t]{0.48\textwidth}
\centering
\includegraphics[
width=\linewidth,
trim={2cm 2cm 2cm 2cm},
clip
]{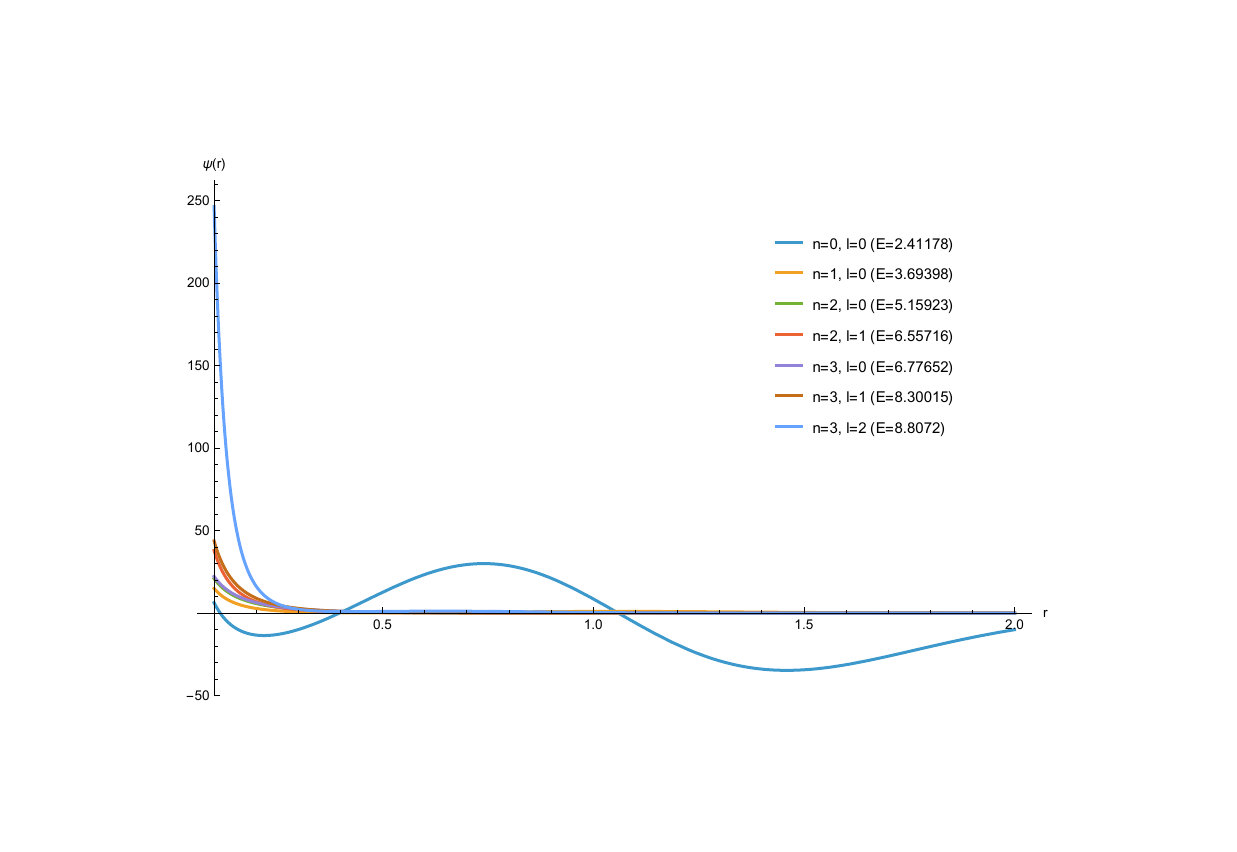}
\caption{Radial wave function in Eq.~(\ref{50})}
\label{fig:cornell50}
\end{subfigure}
\hfill
\begin{subfigure}[t]{0.48\textwidth}
\centering
\includegraphics[
width=\linewidth,
trim={2cm 2cm 2cm 2cm},
clip
]{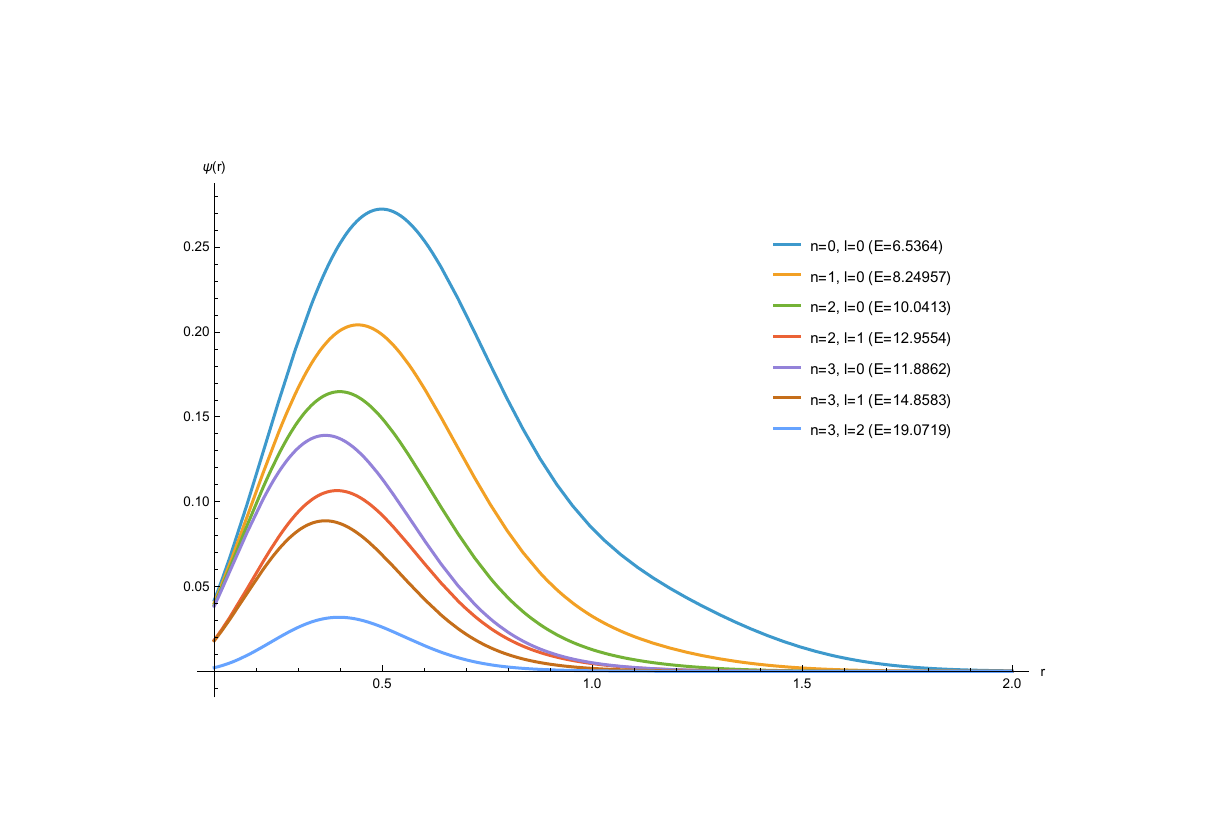}
\caption{Radial wave function in Eq.~(\ref{53})}
\label{fig:cornell53}
\end{subfigure}

\vspace{0.2cm}

\caption{
Radial wave functions involving biconfluent Heun polynomials for the generalized Klein--Gordon oscillator in the Som--Raychaudhuri space--time with $\alpha=0.8$. Panels (a) and (b) correspond to the linear-potential solutions given by Eqs.~(\ref{36}) and (\ref{39}), whereas panels (c) and (d) represent the Cornell-type potential solutions given by Eqs.~(\ref{50}) and (\ref{53}). The plots are shown for different quantum numbers $(n,l)$ using the parameter set $M=e=\Phi_B=\omega=\lambda=B_0=\Omega=k=1$, while $\xi_1=2$ and $\xi_2=1$ are adopted for the Cornell-type potential. The radial wave functions are presented up to an arbitrary normalization constant and are used to illustrate the localization properties and spatial behavior of the bound states.
}
\label{fig:combined}
\end{figure}

\backmatter

%\section*{Declarations}
%\section*{Conflict of Interest}

\section*{Declarations}

\subsection*{Funding}
This work was supported by the KLUBAP-350 Project.

\subsection*{Availability of data and material}
This study is based entirely on analytical calculations. No experimental or observational data are used, and all results necessary to support the conclusions are provided in the manuscript.

\subsection*{Competing interests}
The authors declare no competing interests.

\subsection*{Authors Contributions}
H.K. conceptualized the study, formulated the problem, applied the extended Nikiforov Uvarov method, performed the analytical calculations, interpreted the results, and wrote the original manuscript. T.Ç. contributed to the analytical solution process, performed the numerical calculations, and prepared the figures. D.D. contributed to the interpretation of the results, supervised the study, and critically reviewed and revised the manuscript. All authors read and approved the final manuscript.



\begin{thebibliography}{99}
\bibitem{omar}
O. Mustafa, PDM Klein--Gordon particles in Gödel-type Som--Raychaudhuri cosmic string spacetime background. Eur. Phys. J. Plus 138, 21 (2023)

\bibitem{bruce}
S. Bruce, P. Minning, The Klein--Gordon oscillator. Nuovo Cimento A 106, 711 (1993)

\bibitem{soares}
A.R. Soares, R.L.L. Vitória, H. Aounallah, On the Klein--Gordon oscillator in topologically charged Ellis--Bronnikov-type wormhole spacetime. Eur. Phys. J. Plus 136, 966 (2021)

\bibitem{leite}
E.V.B. Leite, H. Belich, R.L.L. Vitória, Effects of the Cornell-type potential on a position-dependent mass system in Kaluza--Klein theory. Adv. High Energy Phys. 2019, 6740360 (2019)

\bibitem{Bouzenada}
A. Bouzenada, A. Boumali, R.L.L. Vitória et al., Dynamics of a Klein--Gordon oscillator in the presence of a cosmic string in the Som--Raychaudhuri space--time. Theor. Math. Phys. 221, 2193--2206 (2024)

\bibitem{bakke1}
K. Bakke, C. Furtado, On the Klein–Gordon oscillator subject to a Coulomb-type potential. Ann. Phys. 355, 48 (2015)

\bibitem{ahmed1}
F. Ahmed, The generalized Klein–Gordon oscillator in the background of cosmic string space-time with a linear potential in the Kaluza–Klein theory. Eur. Phys. J. C 78, 598 (2020)

\bibitem{ahmed2}
F. Ahmed, Quantum effects on Klein--Gordon oscillator under a Cornell-type potential in Kaluza--Klein theory. Grav. Cosmol. 27, 292--301 (2021)

\bibitem{olivera}
M.D. de Oliveira, A.G.M. Schmidt, Modified Klein--Gordon oscillator in Ellis--Bronnikov-type wormhole spacetime with cosmic string and global monopole. Phys. Scr. 100, 025304 (2025)

\bibitem{zhong}
L. Zhong, H. Chen, Z.W. Long, C.Y. Long, H. Hassanabadi, The study of the generalized Klein--Gordon oscillator in the context of the Som--Raychaudhuri space--time. Int. J. Mod. Phys. A 36, 2150129 (2021)

\bibitem{araujo}
J.C.B. Araújo, J.E.G. Silva, D.F.S. Veras, C.A.S. Almeida, A smoothed string-like braneworld in six dimensions. Eur. Phys. J. C 75, 127 (2015)

\bibitem{hale2}
H. Karayer, D. Demirhan, F. Buyukkilic, Extension of Nikiforov--Uvarov method for the solution of Heun equation. J. Math. Phys. 56, 063504 (2015)

\bibitem{heun} A. Ronveaux, {Heun's Differential Equations} Oxford University Press, New York  (1995)

\bibitem{Nikiforov} A. V. Nikiforov  and  V. B. Uvarov,  { Special Functions of Mathematical Physics}, Birkhauser, Boston (1988)
%\emph{Special Functions of Mathematical Physics}, Birkhauser, Boston, (1988).

\bibitem{berkdemir} M.R. Pahlavani,  {Theoretical Concepts of Quantum Mechanics}, Rijeka, Croatia (2012)



\bibitem{mirza}
B. Mirza, M. Mohadesi,  The Klein-Gordon and the Dirac Oscillators in a Noncommutative Space. Commun. Theor. Phys. 42, 664 (2004)

\bibitem{lutfu}
B.C. Lütfüoğlu, J. Kříž, P. Sedaghatnia et al., The generalized Klein--Gordon oscillator in a cosmic space-time with a space-like dislocation and the Aharonov--Bohm effect. Eur. Phys. J. Plus 135, 691 (2020)



\bibitem{hale3}
H. Karayer, D. Demirhan, F. Buyukkilic, Some special solutions of biconfluent and triconfluent Heun equations in elementary functions by extended Nikiforov--Uvarov method. Rep. Math. Phys. 76, 271--281 (2015)

\bibitem{hale4}
H. Karayer, D. Demirhan, F. Buyukkilic, Solution of Schrödinger equation for two different potentials using extended Nikiforov-Uvarov method and polynomial solutions of biconfluent Heun equation. J. Math. Phys. 59, 053501 (2018)

\bibitem{hale5}
H. Karayer, D. Demirhan, Analytical eigenstate solutions of Schrödinger equation with noncentral generalized oscillator potential by extended Nikiforov-Uvarov method. Phys. Lett. A 413, 127608 (2021)

\bibitem{caruso}
F. Caruso, J. Martins, V. Oguri, Solving a two-electron quantum dot model in terms of polynomial solutions of a biconfluent Heun equation. Ann. Phys. 347, 130--140 (2014)

\bibitem{ishkhanyan}
T.A. Ishkhanyan, A.M. Ishkhanyan, Solutions of the bi-confluent Heun equation in terms of the Hermite functions. Ann. Phys. 383, 79--91 (2017)

\bibitem{aharonov}
Y. Aharonov, D. Bohm, Significance of electromagnetic potentials in the quantum theory. Phys. Rev. 115, 485 (1959)

\bibitem{chen}
H. Chen, Z.W. Long, Q.K. Ran, Y. Yang, C.Y. Long, Relativistic quantum effects associated with the Aharonov--Bohm phase in Gödel-type space-times. EPL 132, 50006 (2020)












\end{thebibliography}
\end{document}